# Polarization shaping of high-order harmonics in laser-aligned molecules

E. Skantzakis[1], S. Chatziathansiou[1,2], P. A. Carpeggiani[5], G. Sansone[3,4,5], A. Nayak[3], D. Gray[1], P. Tzallas[1,3], D. Charalambidis[1,2,3,*], E. Hertz[6], and O. Faucher[3,6,*]

[1] Foundation for Research and Technology-Hellas, Institute of Electronic Structure and Laser, P.O. Box 1527, GR-711 10 Heraklion, Crete, Greece
[2] Department of Physics, University of Crete, P.O. Box 2208, GR71003 Heraklion, Crete, Greece
[3] ELI-ALPS, ELI-Hu Kft., Dugonics tér 13, H-6720 Szeged Hungary
[4] Institute of Photonics and Nanotechnologies (IFN)-Consiglio Nazionale delle Ricerche (CNR), Piazza Leonardo da Vinci 32, 20133 Milano, Italy
[5] Dipartimento di Fisica Politecnico, Piazza Leonardo da Vinci 32, 20133 Milano, Italy
[6] Laboratoire Interdisciplinaire CARNOT de Bourgogne, UMR 6303 CNRS-Université Bourgogne Franche-Comté, 9 Av. A. Savary, BP 47870, F-21078 DIJON Cedex, France
[*] Correspondence and requests for materials should be addressed to O.F. (email: olivier.faucher@u-bourgogne.fr) or D.C. (email: chara@iesl.forth.gr)

## ABSTRACT

The present work reports on the generation of short-pulse coherent extreme ultraviolet radiation of controlled polarization. The proposed strategy is based on high-order harmonics generated in pre-aligned molecules. Field-free molecular alignment produced by a short linearly-polarized infrared laser pulse is used to break the isotropy of a gas medium. Driving the aligned molecules by a circularly-polarized infrared pulse allows to transfer the anisotropy of the medium to the polarization of the generated harmonic light. The ellipticity of the latter is controlled by adjusting the angular distribution of the molecules at the time they interact with the driving pulse. Extreme ultraviolet radiation produced with high degree of ellipticity (close to circular) is demonstrated.

## Introduction

High-harmonic generation (HHG) by intense femtosecond laser pulses has been extensively investigated for the last two decades[1–3]. The research accomplished in this field has led to the emergence of table-top sources of coherent bright, ultrashort extreme ultraviolet (EUV) and Soft X-ray radiation and attosecond pulses[4]. The unique characteristics of these sources make them suitable for a wide range of applications, including time-resolved ultrafast dynamics in gas phase and condensed matter[4]. Most of the studies and applications of HHG have been performed with linearly-polarized HHG, although there is a strong demand for providing harmonics with controllable polarization-state. In particular, circularly-polarized short EUV pulses are very useful for investigating circular dichroism, ultrafast spin dynamics, magnetic microscopy, chirality assignment, and so forth. Several theoretical methods have been proposed for generating circularly-polarized harmonics (see e.g. Refs.[5–7]). However, experimental demonstrations were still very limited until a couple of years ago and in all attempts the harmonic radiation were limited to low photon flux and/or poor ellipticity[8–11]. The past two years have seen the emergence of noteworthy achievements that have overcome the previous limitations. This breakthrough was initiated by the work of Fleisher *et al.*[12], where tunable polarization high-harmonics were generated using coplanar, elliptically-polarized counter-rotating dichroic drivers. Based on a similar strategy, Kfir *et al.*[13] were able to produce circularly-polarized EUV light using circularly-polarized, instead of elliptically-polarized, counter-rotating dichroic drivers and provided the first application of EUV table-top source to the measurement of magnetic circular dichroism (MCD) in thin CO foil enabling in the same way to determine the helicity of the produced circular EUV beam. Hickstein *et al.*[14,15] pushed further the concept of counter-rotating circularly-polarized laser pulses by showing that a non collinear geometry allows, among other advantages, the production of spatially separated circularly-polarized harmonics of the same frequency and opposite helicity. In the mean time, photoelectron circular dichroism of chiral molecules using elliptically-polarized light combined with a resonant interaction was reported by Ferré *et al.*[16] and elliptically-polarized EUV light produced by crossed polarized phase quadrature dichroic fields allowed the measurement of the MCD effect of nickel by Lambert *et al.*[17].

Following a different approach, we have shown that circularly-polarized light can be used to drive harmonic generation in aligned molecules[18]. Among other potential applications, we presented results on controlling the polarization of the generated harmonic radiation. However, the analytical model presented in Ref.[18] was developed in the framework of perturbation theory and the experiment was carried out by detecting the third-harmonic radiation generated with a weak laser field. The purpose of the present letter is to demonstrate that aligned molecules driven by quasi circularly-polarized light, *i.e.*, with an ellipticity set close to unity, can be used to produce elliptically- and circularly-polarized higher-order harmonics generated in the strong field regime. The strategy presented here allows extensive control over the polarization of the generated radiation, complementing the existing toolbox of polarization engineering of high harmonic EUV sources.

## Results

### Production and analyzis of the harmonics

The experiment is based on a 10 Hz amplified Ti:sapphire femtosecond laser system delivering 30 fs pulse duration at $\lambda_0$ =800 nm with a maximum energy of 400 mJ. The whole experimental setup is designed so as to be operated under vacuum conditions. As depicted in Fig. 1, the radiation of a Ti:sapphire fs laser system is split in two so as to provide a sequence of aligning and driving pulse with an adjustable time delay. The aligning pulse is linearly polarized, whereas the second pulse driving the harmonic process can be circularly polarized using a combination of a half- (HWP1) and a quarter-wave ($\lambda/4$) plate. The two beams are focused by a 3-meter focal length and crossed at a small angle into a pulsed jet of carbon dioxide ($CO_2$) gas where the EUV radiations are generated. The intensity of the IR light on the spectrometer is reduced by a silicon wafer working with an incidence angle of 75°, *i.e.*, close to the Brewster angle for 800 nm. The harmonic radiation is then reflected by a silver mirror, dispersed by an EUV spectrometer, and finally imaged on microchannel plates (MCP) detector. The back side of the MCP detector is mounted on a phosphor screen enabling recording the harmonic spectrum with a CCD camera. The polarization of the harmonics is measured thanks to the combination of polarization sensitive optical elements including the Si wafer, the silver mirror, and the grating of the spectrometer, the whole acting as a EUV reflective polarization analyzer favouring *s*-polarized waves. Since the HHG polarization measurements can not be achieved by rotating the analyzer, the former is conducted by rotating both the aligning and the driving pulse around the optical axis using a large-aperture half-wave plate (HWP2) through which both pulses are passing. In the results shown below, the energy of the aligning (driving) pulse is set to 10 (15 mJ) mJ leading to an estimated peak intensity at focus of $\approx 7 \times 10^{13}$ W/cm$^2$ ($\approx 1 \times 10^{14}$ W/cm$^2$).

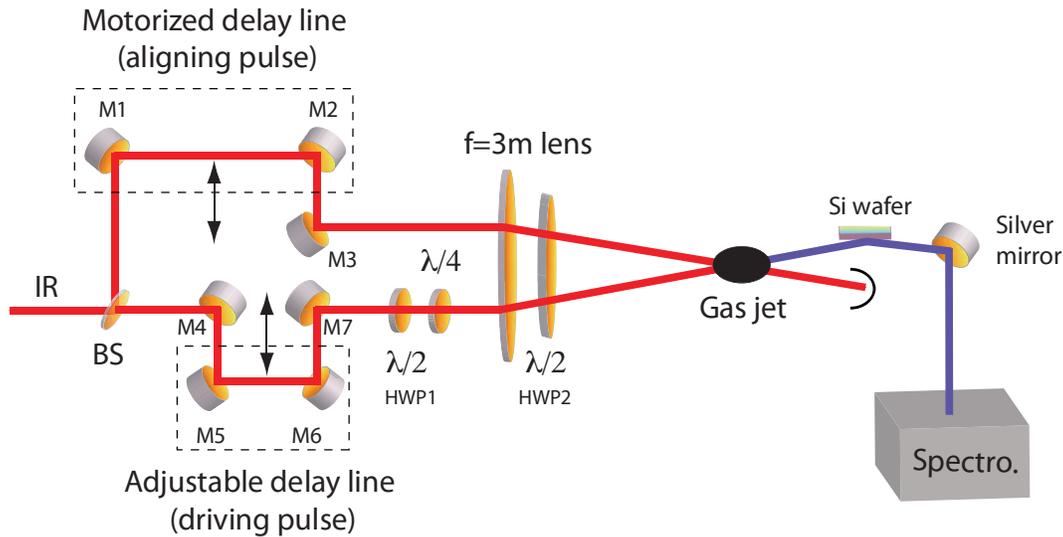

**Figure 1.** Experimental setup for polarization shaping of high-harmonic radiations generated in transiently aligned molecules. M: Mirrors; BS: Beam splitter; HWP: Half-wave plates; $\lambda/4$: Quarter-wave plate.

The characterization of the EUV analyzer is performed by recording the harmonic signal produced by randomly oriented $CO_2$ molecules exposed to a linearly *p*-polarized pulse. For symmetry reasons, the harmonics share the same polarization with the driving field. Although all angular distributions presented in this work have been recorded by rotating the field to be analyzed with the aid of HWP2, for the sake of simplicity we will consider hereafter that the field is fixed whereas the analyzer is rotated. The latter is not standard in so far as its extinction ratio *R*, defined here as the ratio between the detected signals for a linearly-polarized field oriented vertically (*i.e.*, *s*-polarized) and horizontally (*i.e.*, *p*-polarized), is much smaller than what is currently obtained with traditional polarizers. Figure 2 (a) depicts the signal of the 9th harmonic recorded over several turns of HWP2. The angle $\Phi$ describes the orientation of the analyzer axis, defined here as the direction along which its efficiency is maximum, with respect to the horizontal axis. The same signal depicted in a polar coordinates system is shown in Fig. 2 (b). As shown, the horizontally polarized field leads to maxima when $\Phi = 0 \, [\pi]$. An accurate determination of the contrast of the analyser can be deduced from the modulation depth of the signal using a modified Malus' Law whose details can be found in the Supplemental information online. The result of the simulation performed with the fitted contrast value $R = 4.5$ is shown for comparison in the same figures.

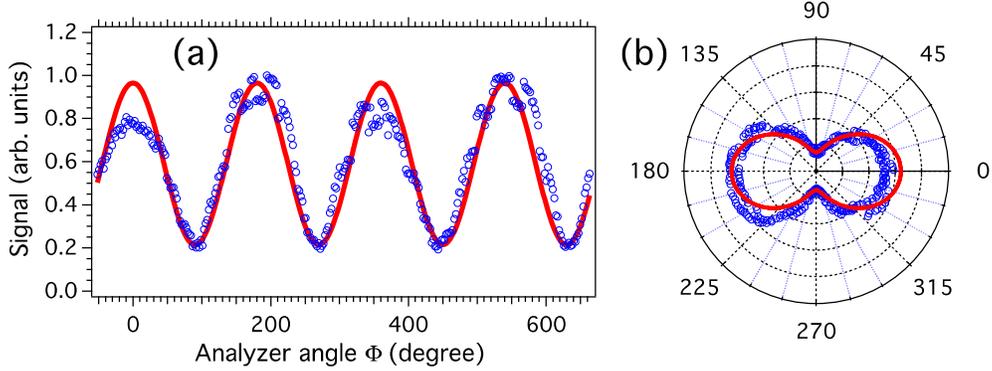

**Figure 2.** (a) Polarization analysis of the linearly-polarized harmonic field generated at 88.9 nm. Each data point represents an averaging over 10 laser shots. (b) The corresponding signal depicted in a polar plot. Experimental (numerical) data are shown with blue circles (red lines).

## Polarization control of high-order harmonics

Next, we consider the possibility to control the polarization of the generated harmonics. Before to address this matter, it should be mentioned that harmonic generation of circularly-polarized radiation is allowed in a gas of aligned molecules. For reviews on laser-induced molecular alignment, see e.g., Refs. [19,20]. This has been demonstrated in the specific case of third-harmonic generation[18]. In a gas of atoms or randomly oriented molecules, the harmonic generation process is forbidden due to the conservation of the spin angular momentum in "atom+field" or "molecule+field" systems presenting a symmetry axis. In an ensemble of aligned molecules, the axial symmetry of the system can be broken if the propagation axis of the circularly-polarized field and the alignment axis are different. In that case, the projection $m$ of the angular momentum $J$ along any axis is not longer a good quantum number. As a consequence, third-harmonic generation and higher-order frequency up conversion processes are allowed. Figure 3 depicts for instance the 9th harmonic generated by a circularly-polarized IR pulse in $CO_2$ molecules for different temporal delays between the aligning and driving pulse. So far, this is the first report of HHG produced with a circularly-polarized field. The alignment field is linearly polarized within the polarization plane of the driving field. The adjustment of the half-wave plate HWP1 with respect to the neutral axis of the quarter-wave plate (see Fig. 1) is performed so as to produce the extinction of the harmonic signal in the absence of the aligning pulse.

As a result of the anisotropy introduced by aligning the molecules, the harmonics generated by the circularly-polarized pulse should be elliptically polarized, the ellipticity of them being governed by the time delay between the alignment pulse and the driving field. In order to verify this assumption, both polarization components of the harmonic field must be resolved. Figure 4 (a) compares the harmonic signal recorded with two orthogonal orientation of the analyzer, namely vertical and horizontal. The difference between the two curves reveals that the polarization of the harmonic field varies with the delay as a result of the change occurring in the angular distribution of the molecules. In order to quantify the impact of these change, the polarization of the harmonic field is measured using the same methods as in Fig. 2. The result is shown in Fig. 4 (b) for a delay set at 11.2 ps. Using the contrast value $R$ determined before, the numerical model allows to estimate the ratio $r = \frac{|E_z|}{|E_y|}$, with $E_y$ ($E_z$) the field component along the horizontal $y$-axis (vertical $z$-axis), and $\phi = \phi_y - \phi_z$ the dephasing between these components. The fitted value of the field component ratio ($r = 0.14$) reveals that the polarization of the field is closed to linear with an orientation along the $y$-axis (*i.e.*, the *p*-wave is dominant). Because of the limited extinction ratio of the EUV analyzer, this small ellipticity produces a signal that is about 3 times larger [see Fig. 4 (a)] when the axis of the analyzer is oriented along the horizontal direction of the field as compared to the vertical one. For comparison, an ideal analyzer would lead with the same field ellipticity to a contrast of about 50. The fitted value of the dephasing is $\phi = 1.33$ rad. The small deviation from the expected value $\pi/2$, which remains within the experimental uncertainty, is responsible for a slight tilt of the angular distribution observed in the fitted curve of Fig. 4 (b). From the determination of $r$ and $\phi$, the ellipticity of the

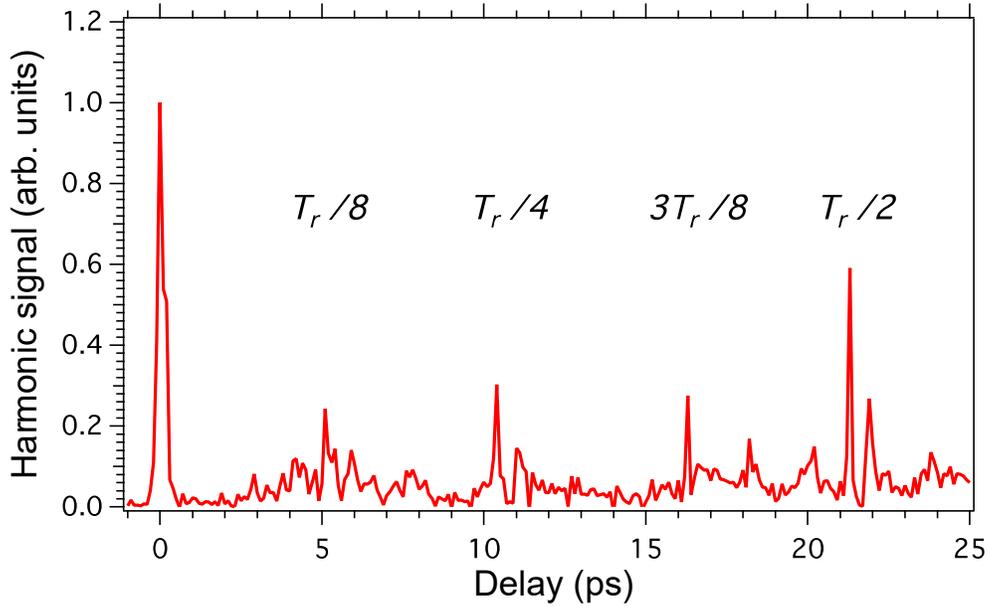

**Figure 3.** 9th harmonic versus delay between a linearly-polarized aligning and a circularly-polarized driving pulse. $T_r = 42.7$ ps: full revival period of $CO_2$.

EUV field defined as $\varepsilon = \tan\left(1/2 \arcsin\left[2r/(1+r^2)\sin\phi\right]\right)$ is estimated to 0.14. As a general remark, it should be pointed out that $\varepsilon$ provides an upper value of the ellipticity since our detection scheme does not allow to differentiate polarized from unpolarized light. Some unpolarized EUV component may be present due to the radial and temporal intensity variation of the lasers. However, due to the high nonlinearity of the harmonic generation process the most efficient generation results from the highest laser intensity part, where the intensity distribution does not vary significantly and thus the unpolarized EUV component remains small.

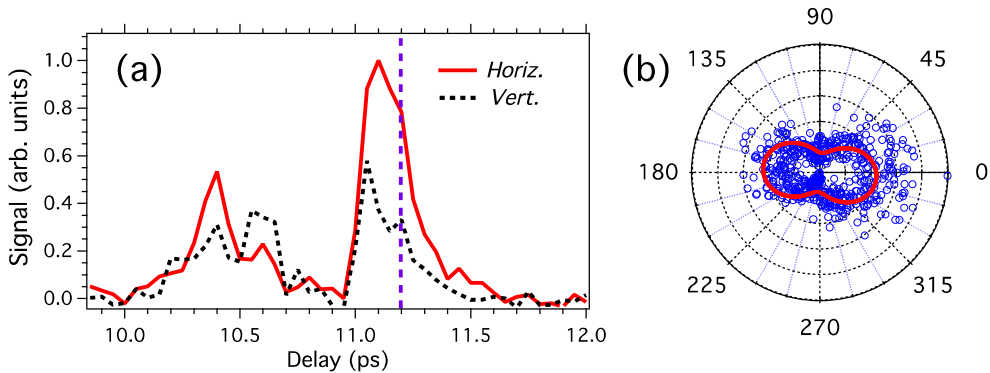

**Figure 4.** (a) 9th harmonic intensity recorded for the horizontal and vertical orientation of the analyzer axis. The dotted vertical line indicates the delay at which the polarization analysis is conducted (see right panel). (b) Polarization analysis with the experimental (numerical) data depicted with blue circles (red line). The fitted ellipticity is $\varepsilon = 0.14$.

It is well known that the phase quadrature between the two oscillating field components is a first prerequisite in achieving a circular polarization. The second one is that the components should share the same amplitude. In principle, the latter can be obtained by adjusting the time delay so as to find a temporal domain where the two signals depicted in Fig. 4 (a) are equal[18]. However, depending on the harmonic order and the degree of alignment, the two field components are not necessarily balanced when the harmonic amplitude is significant. To circumvent this problem, a small ellipticity is introduced along the horizontal or vertical direction of the driving field in order to increase the generation along the corresponding direction. This can be understood as analogous to the addition of a small linear component to a circularly-polarized IR field, the former producing harmonics preferentially along its direction. This approach is applied to the result presented in Fig. 5 (a) where the half-wave plate HWP1 has been slightly tilted from the direction optimizing the suppression of the harmonic signal. The corresponding polarization analysis performed at a delay of 11.2 ps is shown in Fig. 5 (b). The ellipticity and phase value derived from the least square fit are $\varepsilon = 0.85$ and $\phi = 1.73$ rad, respectively. Although the amplitude ratio between the two field components is very close to unity ($r = 0.97$), the small difference between the fitted phase and $\pi/2$ (+9°) generates a small ellipticity of the theoretical curve. The polarization of Fig. 5 (c) is obtained after a small shift of the delay, 11.18 ps and with a different adjustment of HWP1 so as to produce an elliptic polarization. Compared to the two previous cases, the major axis of the ellipse is now standing along the vertical direction, $r = 1.23$, with a phase $\phi = 1.5$ rad (-4°) that remains about constant, leading to an ellipticity $\varepsilon = 0.80$. The polarization shaping is also applicable for other harmonics. For instance, the result of Fig. 5 (d) is obtained for the 7th harmonic with a delay of 10.45 ps. The adjusted ellipticity and phase are $\varepsilon = 0.74$ and $\phi = 1.8$ rad (+13°), respectively. For technical reasons, it was not possible to measure both 7 and 9th harmonic simultaneously. The photon flux of the highly elliptic EUV beam just after the HHG medium is $\approx 4$ pJ, *i.e.*, $2.10^6$ photons/pulse. This is measured by means of an EUV calibrated photodiode in combination with the photoelectron spectra recorded for linear and circular EUV radiation (see the Supplemental information).

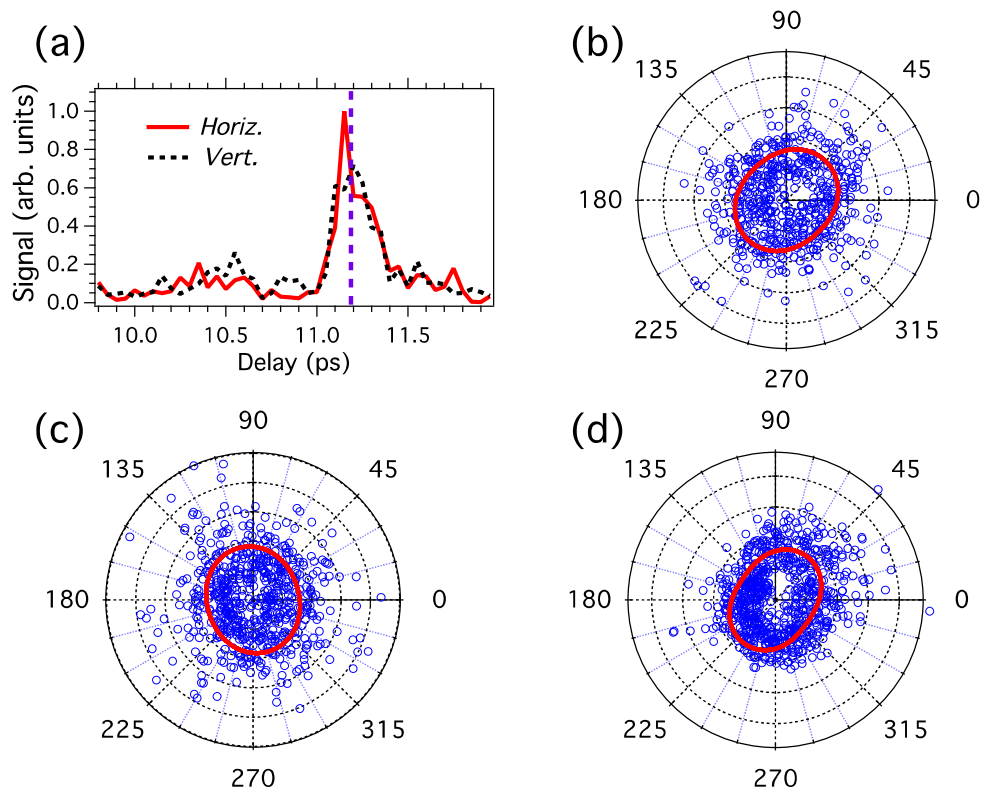

**Figure 5.** (a) Harmonic intensity recorded for the horizontal and vertical orientation of the analyzer axis. Harmonics generated with different polarizations states: (b) 9th harmonic ($\varepsilon = 0.85$), (c) 9th harmonic ($\varepsilon = 0.8$), and (d) 7th harmonic ($\varepsilon = 0.74$) (see text).

## Discussion

We have shown that coherent short pulse EUV radiations with adjustable polarization state, ranging between linear and near circular, can be produced by molecular alignment assisted high-harmonic generation. The effect demonstrated here for high-order harmonics of low photon energies, which are suitable for probing molecular structures below their ionization thresholds, is in principle applicable to high-order harmonics of higher photon energies. The polarization shaping is based on the difference between the field vector components of the generated harmonic, whose relative amplitudes are ruled by the angular distribution of the molecules at the time they interact with a circularly-polarized driving field. Changing the delay between the aligning and driving pulse enables to modify to a large extent the amplitude ratio between the two field projections. In addition, the ability to generate circular polarization requires that the two electric field components are produced in phase quadrature. This condition is fulfilled in the case of third-harmonic generation where it has been shown experimentally and theoretically (in the framework of perturbation theory) that the transverse dipole components oscillate in phase quadrature[18]. One might question whether this also applies to higher-order harmonics. Hereby it is revealed that the high-harmonic field components are in fact produced with the proper phase relation up to the 9th harmonic. It should be noticed that a small rotation of the polarization of the harmonics was reported for atoms driven by polarized light of moderate ellipticity ($0 < \varepsilon < 0.4$)[21]. In the present work, where the ellipticity of the driving field is one (or close to one) the amplitude of the two electric components of the fundamental field are comparable. Therefore, any dephasing occurring between the generated harmonic components, leading to a rotation of the harmonic field ellipse, would most probably result from the anisotropy of the potential of the aligned molecules rather than from the different amplitude of the two driving field components[21].

The harmonics generated with large ellipticity are about two-order of magnitude less intense than the same produced with a linear polarization (see Supplementary Fig. S1 online). This is well understood within the so-called three-step model[22,23] interpreting high-order harmonic generation within a semiclassical approach; compared with a linear polarization, an electron accelerated in an elliptically-polarized field has less chance to be driven back to its parent ion and thus to undergo an inelastic recollision converting its kinetic energy into the emission of a EUV photon. In systems with central symmetry, like atoms, the three-step model prohibits any harmonic generation from circularly-polarized light in agreement with the conservation of the angular momentum. In laser aligned molecules, the process is allowed[24,25] thanks to the anisotropy of the potential. Likewise, the conservation of the angular momentum is fulfilled when considering the total external field applied to the molecule. This last point can be evidenced by describing the harmonic generation process as a multiphoton process. A photon carries a spin angular moment that is $+\hbar$ or $-\hbar$, depending if the circular polarization of the field is left- ($\sigma+$) or right-handed ($\sigma-$), respectively. Figure 6 illustrates the generation of the 9th harmonic by aligned molecules driven by left-handed fundamental photons. The quantification axis is chosen along the propagation direction of the fields, *i.e.*, perpendicular to the aligning and driving fields. In the absence of aligning field the molecules are randomly oriented. The absorption of 9 photons, each transition obeying to the selection rule $\Delta J = \pm 1$, $\Delta m = +1$, leads the system to a virtual state ($J'$, $m+9$) that can not be coupled to the initial state through a single-photon emission, as shown in Fig. 6(a). In this case, the HHG process is forbidden. The action of the aligning pulse is to excite the molecule, by means of intrapulse nonresonant Raman transitions ($\Delta J = 0, \pm 2$, $\Delta m = 0, \pm 2$), to a rotational state ($J'$, $m-8$), with $|m-8| \leqslant J'$[26], as shown in Fig. 6(b). Once the molecule is prepared in this state, it can be driven by the fundamental pulse through a nine-photon transition to a virtual state ($J''$, $m+1$) from which a harmonic photon of same polarization as the fundamental can be emitted, bringing back the system to its initial state. The angular momentum of the "molecule+field" is thus conserved, but with the expense that the 9th harmonic is produced though a 17-photon process involving 8 and 9 photons from the aligning and fundamental field, respectively. Within this multiphoton picture, the difference of EUV yield between the harmonics driven by a linear polarization and those driven by a circular polarization is attributed to the increase of the nonlinearity of the conversion process.

It should be pointed out that harmonics generated by circularly-polarized light in aligned molecules should experience a spectral splitting due to the energy conservation. This can be explained from Figs. 6(b) and (c). In Fig. 6(b), the energy of the 9th harmonic photon differs from the expected value, which is nine times the energy of the fundamental photon. This shift can be interpreted as a manifestation of the rotational Doppler effect observed recently in molecular rotors probed through linear[27–30] and nonlinear[31] optical interactions. A closer inspection to the quantum state excited by the aligning pulse in Fig. 6(b) reveals that the 9th harmonic is generated by molecules rotating in the opposite sense to the driving field, giving rise to a frequency upshift of the generated radiation. In the same time, the aligning pulse also produce counter-rotating molecules relative to the circular polarization of the fundamental field. As shown in Fig. 6(c), those molecules generate downshifted harmonics. Because the linearly-polarized aligning pulse can not orient the angular momentum of the molecule, no preferential sense of molecular rotation is favoured and both of the scenarios depicted in Figs. 6(b) and (c) apply with the same probability. The generated harmonics should be therefore spectrally split in two components. Unfortunately, the resolution of the EUV spectrometer employed in this experiment did not enable to resolve the splitting pf the 7th and 9th harmonic estimated between 6 and 8 THz for the present conditions.

Compared to other existing techniques the present method, utilizing two external control parameters, namely the delay between the two pulses and the ellipticity (close to unity) of the driving EUV field, offers significant control over the polarization of the high-order harmonics generated by a nonresonant single-color driving laser. In contrast to methods relying on the synchronisation of polarized femtosecond driving pulses[12–15,17], the polarization control mechanism investigated here does not require a stabilization of the relative phase between the two laser (aligning and driving) pulses with interferometric accuracy. It is also applicable to all the existing methods of producing isolated attosecond pulses with few or many cycle driving pulses and thus it can supply circularly polarized isolated pulses from the central part (highest intensity) of the driving laser pulse, assuming that the spectral components of the EUV continuum are generated with essentially the same ellipticity.

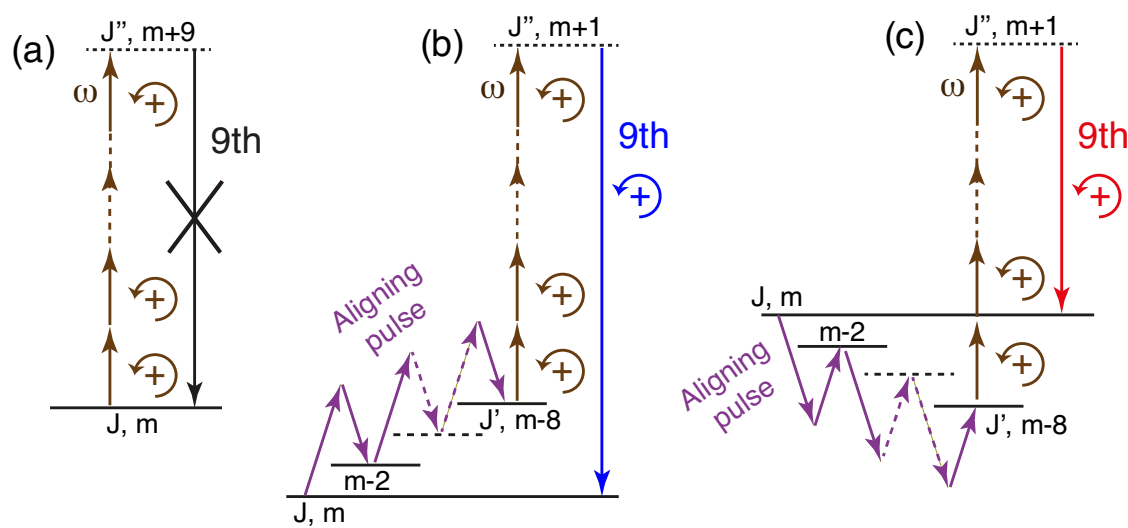

**Figure 6.** Generation of the 9th harmonic depicted in the multiphoton representation. The molecules are aligned with a linearly-polarized pulse and the harmonic is generated by a left-handed circularly-polarized fundamental pulse ($\sigma+$) of frequency $\omega$; (a) only the field driving the harmonic generation is applied; (b) the molecule is first aligned and then exposed to the fundamental driving pulse; (c) the same as (b) but with different intermediate transitions.

While the question regarding the ellipticity variation in subsequent harmonics seeks for further experimental investigations, the ellipticity control of individual harmonics demonstrated here is important for applications in itself. Compared with recent techniques employing counter-rotating circularly-polarized femtosecond laser pulses reporting similar efficiency with standard HHG process[12–14], the moderate efficiency of the present method limits its application to high peak power lasers.

The polarization of the EUV light produced by HHG in molecules depends on the structure of the molecular orbitals, the electronic dynamics in the ion, and the orientation of the molecule with respect to the field[32]. It will be fruitfull to interpret our data at the light of existing theoretical models. This would allow, in one one hand, to assess the impact of circularly or quasi-circularly polarized light on the HHG dynamics of aligned molecules[24,25] and, on the other hand, to extract generic molecular features by decrypting the information encoded in the emitted EUV radiation through a complete analysis of its polarization state[33].

## Acknowledgements

The research leading to these results has received funding from LASERLAB-EUROPE (grant agreement no. 284464, EC's Seventh Framework Programme) and has been further supported in part by the European Union's Horizon 2020 research and innovation programme under the Marie Skłodowska-Curie grant agreement No 641789 MEDEA, the Greek funding program NSRF and the Labex ACTION program (contract ANR-11-LABX-0001-01).

## Author contributions statement

D.C., G.S., and O.F. initiated the project. E.S. and P.T. designed the experiment., E.S., S.C., P.A.C., A.N., E.H., and O.F. conducted the experiments. D.G. operated the laser system. P.T., D.C., E.H., and O.F. supervised the project, and O.F. wrote the manuscript. E.S., G.S., P.T., D.C., E.H., and O.F. analyzed the results and contributed to the manuscript.

## Additional information

**Supplementary information** accompanies this paper at http://www.nature.com/srep
**Competing financial interests:** the authors declare no competing financial interests.

# Polarization shaping of high-order harmonics in laser-aligned molecules: Supplementary

E. Skantzakis[1], S. Chatziathansiou[1,2], P. A. Carpeggiani[5], G. Sansone[3,4,5], A. Nayak[3], D. Gray[1], P. Tzallas[1,3], D. Charalambidis[1,2,3,*], E. Hertz[6], and O. Faucher[3,6,*]

[1] Foundation for Research and Technology-Hellas, Institute of Electronic Structure and Laser, P.O. Box 1527, GR-711 10 Heraklion, Crete, Greece
[2] Department of Physics, University of Crete, P.O. Box 2208, GR71003 Heraklion, Crete, Greece
[3] ELI-ALPS, ELI-Hu Kft., Dugonics tér 13, H-6720 Szeged Hungary
[4] Institute of Photonics and Nanotechnologies (IFN)-Consiglio Nazionale delle Ricerche (CNR), Piazza Leonardo da Vinci 32, 20133 Milano, Italy
[5] Dipartimento di Fisica Politecnico, Piazza Leonardo da Vinci 32, 20133 Milano, Italy
[6] Laboratoire Interdisciplinaire CARNOT de Bourgogne, UMR 6303 CNRS-Université Bourgogne Franche-Comté, 9 Av. A. Savary, BP 47870, F-21078 DIJON Cedex, France
[*] Correspondence and requests for materials should be addressed to O.F. (email: olivier.faucher@u-bourgogne.fr) or D.C. (email: chara@iesl.forth.gr)

## ABSTRACT

Supplementary material for manuscript "Polarization shaping of high-order harmonics in laser-aligned molecules"

## 1 EUV reflective analyzer

The harmonic field polarized in the $Oyz$ plan is written $\vec{E}_H(0,E_y,E_z)$, with $E_y = |E_y|\exp(i\phi/2)$ and $E_z = |E_z|\exp(-i\phi/2)$, where $\phi$ is the dephasing between the two field components. The reflective EUV polarization analyzer consists on a combination of various polarization sensitive optical elements including the silicon wafer, the silver mirror, and the grating of the spectrometer. It induces along the $y$- (i.e., $p$-polarization) and $z$-component (i.e., $s$-polarization) of the electric field a complex amplitude $a_p = |a_p|e^{i\psi_p}$ and $a_s = |a_s|e^{i\psi_s}$, respectively. $\psi_p - \psi_s$ accounts for the eventual dephasing introduced by the analyzer. The action of the analyzer is characterised by the tensor

$$\overleftrightarrow{A} = \begin{pmatrix} 0 & 0 & 0 \\ 0 & a_p e^{i\psi_p} & 0 \\ 0 & 0 & a_s e^{i\psi_s} \end{pmatrix} \quad (1)$$

After reflection on the analyzer, the HHG field writes $E_a = \overleftrightarrow{A}\cdot\vec{E}$. The dependence of its intensity with respect to the orientation $\Phi$ of the analyzer is described by the Malus' law

$$\mathscr{I}_a(\Phi) \propto |a_p|^2 \left[ |E_y|^2 \left(\cos^2\Phi + R\sin^2\Phi\right) + |E_z|^2 \left(R\cos^2\Phi + \sin^2\Phi\right) + |E_y||E_z|(R-1)\sin 2\Phi \cos\phi \right], \quad (2)$$

where $R = \left(\frac{|a_s|}{|a_p|}\right)^2$ is defining the extinction ratio of the analyzer.

The two amplitude components of the incident field can be obtained by measuring the signal for two orientations of the analyzer corresponding to $\Phi = 0$ and $\Phi = \pi/2$:

$$|E_y|^2 \simeq \frac{1}{(1+R)} \frac{\mathscr{I}_a(\Phi=0) - R\mathscr{I}_a(\Phi=\pi/2)}{\mathscr{I}_a(\Phi=0) - \mathscr{I}_a(\Phi=\pi/2)}, \quad (3)$$

$$|E_z|^2 \simeq \frac{1}{(1+R)} \frac{\mathscr{I}_a(\Phi=\pi/2) - R\mathscr{I}_a(\Phi=0)}{\mathscr{I}_a(\Phi=0) - \mathscr{I}_a(\Phi=\pi/2)}. \quad (4)$$

The IR field $\vec{E}_0$ delivered by the laser system is polarized along the horizontal $y$ axis. A half-wave plate (HWP1) oriented at $45°/2$ with respect to the $y$ axis is used to convert the linear into a circular polarization. In order to balance the amplitude of the harmonic between its components $E_y$ and $E_z$, a small ellipticity can be induced along the horizontal or vertical direction of the driving field. This is achieved by tilting HWP1 by a very small angle.

## 2 Energy measurement of the near-circularly polarized 9th harmonic

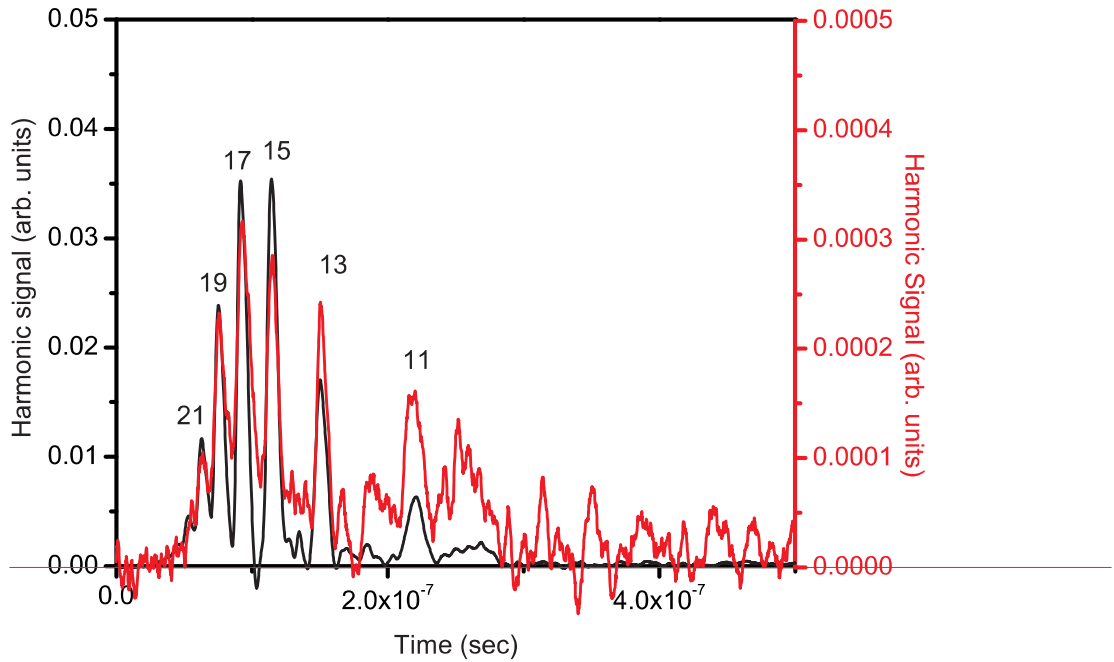

**Figure 1.** Photoelectron (PE) spectra resulted by the interaction of high-order harmonics with Ar (IP=15.76 eV). The black solid line shows the PE spectrum ($S^{(PE)}_{p\text{-pol.}}$) recorded in case of using only the $p$-polarized probe IR beam. In this case the energy of the EUV ($E_{EUV}$) has been measured by means of a calibrated EUV photodiode. The red line shows the PE spectrum ($S^{(PE)}_{c\text{-pol.}}$) recorded in case of using a $p$-polarized probe beam which results to the generation of a $p$-polarized 9th harmonic with an energy approximately equal to the energy of the near-circularly polarized 9th harmonic.

The energy of the near-circularly polarized 9th harmonic has been obtained by means of a calibrated EUV photodiode and a time-of-flight (TOF) photoelectron (PE) spectrometer, placed just after the EUV photodiode. The energy of the high-order harmonics ($E_{EUV}$) generated using only the $p$-polarized probe IR beam has been measured with the calibrated EUV photodiode, placed after the silicon plate and a 150-nm-thick aluminum filter, which transmits harmonics with order $\geqslant$ 11th. This was done by moving the silver mirror (Fig. 1 of the main text of the manuscript) out of the harmonic beam path. The corresponding PE spectrum ($S^{(PE)}_{p\text{-pol.}}$) induced by the interaction of the harmonics with order $\geqslant$ 11th with argon atoms was recorded (black line in Fig. S1) by moving the EUV photodiode out of the harmonic beam path. In this way the correspondence between the PE spectrum and the EUV energy has been established. A PE spectrum ($S^{(PE)}_{c\text{-pol.}}$) has been also recorded (red line in Fig. S1) after reduction of the energy of the 9th harmonic generated by the $p$-polarized IR beam to a value (signal recorded by the EUV spectrometer) equal with the energy of the near-circularly polarized 9th harmonic. This measurement gives the correspondence between the $S^{(PE)}_{c\text{-pol.}}$ spectrum and the energy of the near-circularly polarized 9th harmonic. With the above measurements the energy of the near-circularly polarized 9th harmonic just after the harmonic generation medium has been estimated by using the relation

$$E_{9\text{th}} \approx C_q \frac{E_{\text{EUV}}^{(\text{ph})}}{Q T_{\text{Al}} R_{\text{Si}}} \Big/ \frac{S_{p\text{-pol.}}^{(\text{PE})}}{S_{c\text{-pol.}}^{(\text{PE})}}, \qquad (5)$$

with $Q$ ($\approx 2$), $T_{\text{Al}}$ ($\approx 5$), $R_{\text{Si}}$ ($\approx 0.6$), and $C_q$ ($\approx 0.2$) being the quantum efficiency of the photodiode, the transmission of the Al filter, the reflectivity of the Si plate, and $C_q = S_{p\text{-pol.}}^{(9\text{th})}/S_{p\text{-pol.}}^{(\text{PE})}$ (where $S_{p\text{-pol.}}^{(9\text{th})}$ is the PE signal of the 9th harmonic), respectively. In the above calculation it has been assumed that the energy of the 9th harmonic is approximately equal with the energy of the 11th, 13th harmonics, *i.e.*, $S_{p\text{-pol.}}^{(9\text{th})} = S_{p\text{-pol.}}^{(11\text{th},13\text{th})}$. This approximation is valid since these harmonics are laying in the plateau region of the spectrum.